\documentclass{raa}

\usepackage{graphicx,times}             %for PS/EPS graphics inclusion, new
\usepackage{amsmath}
\usepackage{natbib}
\usepackage{url}
\usepackage{amsmath}
\usepackage{amssymb}

\begin{document}

   \title{Spacecraft Doppler tracking with possible violations of LLI and LPI: a theoretical modeling\,$^*$
\footnotetext{$*$ Supported by the National Natural Science Foundation of China.}
}
   \volnopage{Vol.0 (200x) No.0, 000--000}      %%preserved for Editor. DOn't remove!
   \setcounter{page}{1}          %%starting page, preserved for Editor. DOn't remove!

 \author{Xue-Mei Deng
      \inst{1,3}
   \and Yi Xie
      \inst{2,3}
   }
   %% Here is an example of three authors come from different institutes.
%% For single author or all the authors from an institute, use "\inst{}" only

   \institute{
   Purple Mountain Observatory, Chinese Academy of Sciences, Nanjing 210008, China; \\
%% Please give the E-mail address of the author, to whom future correspondence and
%% offprint requests will be sent.
        \and
             School of Astronomy $\&$ Space Science, Nanjing University, Nanjing 210093, China; {\it yixie@nju.edu.cn}\\
           \and
              Key Laboratory of Modern Astronomy and Astrophysics, Nanjing University, Ministry of Education, Nanjing 210093, China
   }

   \date{Received~~ month day; accepted~~~~month day}

\abstract{
Currently two-way and three-way spacecraft Doppler tracking techniques are widely used and playing important roles in control and navigation for deep space missions. Starting from one-way Doppler model, we extend the models of two-way and three-way Doppler by making them include possible violations of the local Lorentz invariance (LLI) and the local position invariance  (LPI) in order to test the Einstein equivalence principle which is the cornerstone of general relativity and all other metric theories of gravity. After taking the finite speed of light into account, which is so-called light-time solution (LTS), we have these models depending on the time of reception of the signal only for practical convenience. We find that possible violations of LLI and LPI can not affect two-way Doppler under linear approximation of LTS although this approximation is sufficiently good for most cases in the solar system. We also show that, in three-way Doppler, possible violations of LLI and LPI associate with two stations only, which suggests that it is better to set the stations at places with significant differences in velocities and gravitational potentials to obtain high level of sensitivity for tests.
\keywords{space vehicles -- techniques: radial velocities -- gravitation}
}

   \authorrunning{X.-M. Deng \& Y. Xie}            %author_head in even pages
   \titlerunning{Doppler with violations of LLI and LPI}  % title_head in odd pages

   \maketitle
%% The author head (on even pages) and the title head (on odd pages) will be
%% automatically extracted from \author{} and \title{}. Whenever the title is too long,
%% you will be asked to supply a shorter one by inserting either \authorrunning{} or
%% \titlerunning{} before \maketitle. Anyway, you can specify your own heads.
%%
%%
%% Note: In the following text body of your manuscript, please note several differences from
%%       other major journals:
%% (1) \subsection{Please Capitalize the First Letter of Each Notional Word in Subsection Title}
%% (2) Please Capitalize the First Letter of Each Notional Word in all tables' captions

%
%________________________________________________ sections below
%

\section{Introduction}
%% first-level sections will be auto-capitalized
\label{sec:intro}

As one of the currently most important methods for determining the motion of a spacecraft, the Doppler tracking technique has been successfully performed in many deep space missions for control and navigation \citep{Kruger1965NASATMX,Moyer2000}. Besides, it can also be used for a variety of scientific applications, such as fundamental physics. The measurement of the frequency shift in the links connecting \textit{Cassini} spacecraft and the Earth yields a stringent test that proves the validity of general relativity (GR) in the solar system \citep{Bertotti2003Nature425.374}. On the other hand, \citet{Kopeikin2007PLA367.276} point out that this test of GR is under a restrictive condition that the Sun gravitational field is static, and if this restriction is removed the test becomes less stringent. It is also known that Doppler tracking might be the only possible way to detect specific low-frequency ($10^{-5}$ -- 1 Hz) gravitational waves \citep[see][for a recent review]{Armstrong2006LRR9.1}. In this work, we focus on its another application in fundamental physics for testing the Einstein equivalence principle (EEP), which is the ``heart and soul'' of gravitational theory \citep[see][for reviews]{Will1993TEGP,Will2006LRR9.3}. It is worth mentioning that a number of notable scientists, including V. Fock, J. Synge, F. Rohlrich, and others, do \textit{not} support this ``heart and soul'' opinion \citep[see][for a historical review of pro and contra of EEP]{Norton1993RPP56.791}. Although this disagreement may even persist today, we stay with EEP in this work.

EEP is the cornerstone for building GR and all other metric theories of gravity. It states that (1) the trajectory of a freely falling test body is independent on its internal structure and composition, so-called the weak equivalence principle (WEP); (2) the outcome of any local non-gravitational experiment is independent on the velocity of the freely-falling reference frame where it is performed, so-called the local Lorentz invariance (LLI); and (3) the outcome of any local non-gravitational experiment is independent on where and when in the universe it is performed, so-called the local position invariance (LPI) \citep[see][for more details]{Will1993TEGP,Will2006LRR9.3}. The second and third pieces of EEP, i.e. LLI and LPI, can be tested by measuring the frequency of a signal transmitted from a clock as it moves in the gravitational field of a massive body \citep[e.g.][]{Krisher1990MPLA5.1809}. 

Gravity Probe A (GP-A) was launched by NASA in 1976. It carried a hydrogen maser oscillator nearly vertically upward to $10^{7}$ m in the Earth's gravitational field and confirmed that the agreement of the observed relativistic frequency shift with prediction was at the level of $7\times 10^{-5}$ \citep{Vessot1980PRL45.2081}. The \textit{Voyager} flybys of Saturn in 1980 made the first test of an extraterrestrial gravitational redshift and it verified the prediction of EEP to an accuracy of $1\%$ as the spacecraft moved in and out of the gravitational field of Saturn \citep{Krisher1990PRL64.1322}. During flybys of Venus and Earth in 1990, the \textit{Galileo} mission performed the solar redshift experiment and confirmed the total frequency shift predicted by EEP to $0.5\%$ accuracy and the solar gravitational redshift to $1\%$ accuracy \citep{Krisher1993PRL70.2213}.

In these experiments, they all relied on a one-way radio signal transmitted from the spacecraft to the ground stations. The transmitted frequency was referred the onboard clock or frequency standard, while the received signal was referred to these standards at the stations. However, one-way Doppler has practical problems for precision tracking of spacecraft. Onboard frequency standards are significantly less stable than ground-based standards and they are limited by their own noise. One solution for this is to use two-way Doppler tracking. In the two-way mode, the ground station emits a radio signal referenced to a high-quality frequency standard. Then, the spacecraft receives this signal and phase-coherently retransmits it to the earth. The transponding process adds noise, but at negligible levels in current observations and does not require a good oscillator on the spacecraft \citep[see][for a review]{Armstrong2006LRR9.1}. In the two-way Doppler, it forms a close-loop for a signal. In the three-way mode, it has a open-loop that Station 1 emits the signal and the transponded signal is received by Station 2.

Therefore, considering these advantages, we will theoretically extend relativistic models of two-way and three-way Doppler tracking by including possible violations of LLI and LPI as the first step. Case studies will be left in the subsequent works. In Section \ref{sec:wLLILPI}, starting from one-way Doppler, we will construct these models. Since the radio signals travel with finite speed, the light-time solution will be corrected in Section \ref{sec:lighttime}. Conclusions and discussion will be presented in  Section \ref{sec:con}.

\section{Doppler with violations of LLI and LPI}
\label{sec:wLLILPI}

In the following investigation, we will build our models of two-way and three-way Doppler tracking within the solar system barycentric reference system by starting from one-way Doppler \citep{Krisher1993PRL70.2213}.

\subsection{One-Way Doppler}

It is well known that EEP predicts the shift of the frequency \citep{Weinberg1972Book,MTW}. The observed redshift $z$ is defined as
\begin{equation}
  \label{}
  1+z = \frac{\nu_{\mathrm{R}}(t_{\mathrm{R}})}{\nu_{\mathrm{E}}(t_{\mathrm{E}})},
\end{equation}
where $\nu_{\mathrm{E}}(t_{\mathrm{E}})$ is the frequency of an emitted signal at the time $t_{\mathrm{E}}$ and $\nu_{\mathrm{R}}(t_{\mathrm{R}})$ is the frequency of the received signal at the time $t_{\mathrm{R}}$. In the following parts of this work, we will omit these dependence on $t_{\mathrm{E}}$ and $t_{\mathrm{R}}$ in notations so that $\nu_{\mathrm{E}} \equiv \nu_{\mathrm{E}}(t_{\mathrm{E}})$ and $\nu_{\mathrm{R}} \equiv \nu_{\mathrm{E}}(t_{\mathrm{R}})$ unless we specify exceptional cases. Up to the order of $\epsilon^2$ where $\epsilon \equiv c^{-1}$ and $c$ is the speed of light, this relation can be written as \citep{Brumberg1991Book,Krisher1993PRL70.2213,Kopeikin2011Book}
\begin{eqnarray}
  \label{onewaydoppler}
  \frac{\nu_{\mathrm{R}}}{\nu_{\mathrm{E}}} & = & 1 + \epsilon \bm{K}\cdot[\bm{v}_{\mathrm{R}}(t_{\mathrm{R}}) - \bm{v}_{\mathrm{E}}(t_{\mathrm{E}}) ] -\epsilon^2 [\bm{K}\cdot\bm{v}_{\mathrm{R}}(t_{\mathrm{R}})][\bm{K}\cdot\bm{v}_{\mathrm{E}}(t_{\mathrm{E}})] + \epsilon^2 [\bm{K}\cdot\bm{v}_{\mathrm{E}}(t_{\mathrm{E}})]^2\nonumber\\
  & & +\frac{1}{2}\epsilon^2 [\bm{v}^2_{\mathrm{R}}(t_{\mathrm{R}}) - \bm{v}^2_{\mathrm{E}}(t_{\mathrm{E}})] + \epsilon^2 \{U[\bm{y}_{\mathrm{R}}(t_{\mathrm{R}})] - U[\bm{y}_{\mathrm{E}}(t_{\mathrm{E}})]\}\nonumber\\
  & & + \mathcal{O}(\epsilon^3),
\end{eqnarray}
where $\bm{y}_{\mathrm{E}}$ and $\bm{y}_{\mathrm{R}}$ are respectively the positional vectors of the emitter and the receiver, $\bm{v}_{\mathrm{E}}$ and $\bm{v}_{\mathrm{R}}$ are the velocities of them, $t_{\mathrm{E}}$ and $t_{\mathrm{R}}$ are the times of emission and reception, and the unit vector $\bm{K}$ is 
\begin{equation}
  \label{vectorK}
  \bm{K} = - \frac{\bm{y}_{\mathrm{R}}(t_{\mathrm{R}})-\bm{y}_{\mathrm{E}}(t_{\mathrm{E}})}{|\bm{y}_{\mathrm{R}}(t_{\mathrm{R}})-\bm{y}_{\mathrm{E}}(t_{\mathrm{E}})|} .
\end{equation}
Here, $U[\bm{y}_{\mathrm{E}}(t_{\mathrm{E}})]$ and $U[\bm{y}_{\mathrm{R}}(t_{\mathrm{R}})]$ are the Newtonian gravitational potentials at the emitter and the receiver and they can be written as
\begin{equation}
  \label{}
  U[\bm{y}_{\mathrm{R}}(t_{\mathrm{R}})]  = \sum_{A} U_A[\bm{y}_{\mathrm{R}}(t_{\mathrm{R}})] \quad\mathrm{and}\quad U[\bm{y}_{\mathrm{E}}(t_{\mathrm{E}})] = \sum_{A} U_A[\bm{y}_{\mathrm{E}}(t_{\mathrm{E}})].
\end{equation}
In Equation (\ref{onewaydoppler}), all velocity-dependent terms originate in special relativity, while the terms depending on the gravitational potentials are predicted by GR.

In order to test EEP, following \citet{Krisher1993PRL70.2213}, we adopt the parametrization of one-way Doppler [see Equation (\ref{onewaydoppler})] as
\begin{eqnarray}
  \label{onewaydopplerLLILPI}
  \frac{\nu_{\mathrm{R}}}{\nu_{\mathrm{E}}} & = & 1 + \epsilon \bm{K}\cdot[\bm{v}_{\mathrm{R}}(t_{\mathrm{R}}) - \bm{v}_{\mathrm{E}}(t_{\mathrm{E}})] -\epsilon^2 [\bm{K}\cdot\bm{v}_{\mathrm{R}}(t_{\mathrm{R}})][\bm{K}\cdot\bm{v}_{\mathrm{E}}(t_{\mathrm{E}})] + \epsilon^2 [\bm{K}\cdot\bm{v}_{\mathrm{E}}(t_{\mathrm{E}})]^2\nonumber\\
  & & +\frac{1}{2}\epsilon^2 \beta_{\mathrm{R}} \bm{v}^2_{\mathrm{R}}(t_{\mathrm{R}}) - \frac{1}{2}\epsilon^2 \beta_{\mathrm{E}} \bm{v}^2_{\mathrm{E}}(t_{\mathrm{E}}) + \epsilon^2 \sum_{A} \alpha_{\mathrm{R}}^{A} U_A[\bm{y}_{\mathrm{R}}(t_{\mathrm{R}})] -  \epsilon^2 \sum_{A} \alpha_{\mathrm{E}}^{A} U_A[\bm{y}_{\mathrm{E}}(t_{\mathrm{E}})] \nonumber\\
  & & + \mathcal{O}(\epsilon^3).
\end{eqnarray}
Equation (\ref{onewaydopplerLLILPI}) describes the shift of the frequency with possible violations of LLI and LPI. Here, violations of LLI can be tested by fitting the dimensionless parameters $\beta_{\mathrm{R}}$ and $\beta_{\mathrm{E}}$. If LLI is valid, then $\beta_{\mathrm{R/E}}=1$. Violations of LPI can be tested by fitting the dimensionless parameters $\alpha_{\mathrm{R}}^{A}$ and  $\alpha_{\mathrm{E}}^{A}$. If LPI holds true, $\alpha_{\mathrm{R/E}}^{A}=1$. 

For separating these possible violations, we will also use notations $\bar{\beta}_{\mathrm{R/E}}\equiv \beta_{\mathrm{R/E}}-1$ and $\bar{\alpha}^A_{\mathrm{R/E}} \equiv \alpha^A_{\mathrm{R/E}}-1$. Equation (\ref{onewaydopplerLLILPI}) can be rewritten as
\begin{equation}
  \label{onewaydopplerLLILPIg}
  \frac{\nu_{\mathrm{R}}}{\nu_{\mathrm{E}}}\bigg|_{\mathrm{E\rightarrow R}} \equiv \mathcal{F}_{\mathrm{E\rightarrow R}}(t_{\mathrm{E}},t_{\mathrm{R}})  =  \hat{\mathcal{F}}_{\mathrm{E\rightarrow R}}(t_{\mathrm{E}},t_{\mathrm{R}}) + \bar{\mathcal{F}}_{\mathrm{E\rightarrow R}}(t_{\mathrm{E}},t_{\mathrm{R}}) + \mathcal{O}(\epsilon^3),
\end{equation}
where $\hat{\mathcal{F}}_{\mathrm{E\rightarrow R}}(t_{\mathrm{E}},t_{\mathrm{R}})$ represents the shift of the frequency as predicted by GR as
\begin{eqnarray}
  \label{}
  \hat{\mathcal{F}}_{\mathrm{E\rightarrow R}}(t_{\mathrm{E}},t_{\mathrm{R}}) & =  & 1 + \epsilon \bm{K}\cdot[\bm{v}_{\mathrm{R}}(t_{\mathrm{R}}) - \bm{v}_{\mathrm{E}}(t_{\mathrm{E}})] -\epsilon^2 [\bm{K}\cdot\bm{v}_{\mathrm{R}}(t_{\mathrm{R}})][\bm{K}\cdot\bm{v}_{\mathrm{E}}(t_{\mathrm{E}})] + \epsilon^2 [\bm{K}\cdot\bm{v}_{\mathrm{E}}(t_{\mathrm{E}})]^2\nonumber\\
  & & +\frac{1}{2}\epsilon^2  \bm{v}^2_{\mathrm{R}}(t_{\mathrm{R}}) - \frac{1}{2}\epsilon^2 \bm{v}^2_{\mathrm{E}}(t_{\mathrm{E}}) + \epsilon^2 \sum_{A}  U_A[\bm{y}_{\mathrm{R}}(t_{\mathrm{R}})] -  \epsilon^2 \sum_{A}  U_A[\bm{y}_{\mathrm{E}}(t_{\mathrm{E}})],
\end{eqnarray}
and $\bar{\mathcal{F}}_{\mathrm{E\rightarrow R}}(t_{\mathrm{E}},t_{\mathrm{R}})$ indicates the effects caused by possible violations of LLI and LPI as
\begin{equation}
  \label{}
  \bar{\mathcal{F}}_{\mathrm{E\rightarrow R}}(t_{\mathrm{E}},t_{\mathrm{R}})  =  \frac{1}{2}\epsilon^2 \bar{\beta}_{\mathrm{R}}  \bm{v}^2_{\mathrm{R}}(t_{\mathrm{R}}) - \frac{1}{2}\epsilon^2 \bar{\beta}_{\mathrm{E}} \bm{v}^2_{\mathrm{E}}(t_{\mathrm{E}}) + \epsilon^2 \sum_{A} \bar{\alpha}^A_{\mathrm{R}} U_A[\bm{y}_{\mathrm{R}}(t_{\mathrm{R}})] -  \epsilon^2 \sum_{A} \bar{\alpha}^A_{\mathrm{E}} U_A[\bm{y}_{\mathrm{E}}(t_{\mathrm{E}})].
\end{equation}
Equation (\ref{onewaydopplerLLILPIg}) will be used to model two-way and three-way Doppler.

\subsection{Two-Way Doppler}

In the two-way Doppler tracking, a ground station (S) emits a radio signal $\nu_{\mathrm{E}}$ at time $t_{\mathrm{E}}$ and a spacecraft (P) receives the signal with the frequency $\nu'$ at time $t'$; then, the spacecraft (P) transmits the radio signal $q\nu'$ back immediately where $q$ is a known ratio between two integers; and the station (S) receives the signal with the frequency $\nu_{\mathrm{R}}$ at time $t_{\mathrm{R}}$. The whole procedure can be decompose as two one-way Doppler and the shift of the frequency in this close-loop can be easily and concisely expressed as
\begin{equation}
  \label{twowaydopplerLLILPIg}
  \frac{\nu_{\mathrm{R}}}{q\nu_{\mathrm{E}}}\bigg|_{\mathrm{S\rightarrow P \rightarrow S}} = \frac{\nu'}{\nu_{\mathrm{E}}} \cdot \frac{\nu_{\mathrm{R}}}{q\nu'} = \mathcal{F}_{\mathrm{S\rightarrow P}}(t_{\mathrm{E}},t') \cdot \mathcal{F}_{\mathrm{P\rightarrow S}}(t',t_{\mathrm{R}}) + \mathcal{O}(\epsilon^3),
\end{equation}
whose explicit form can be written as
\begin{eqnarray}
  \label{twowaydopplerLLILPIex}
  \frac{\nu_{\mathrm{R}}}{q\nu_{\mathrm{E}}} \bigg|_{\mathrm{S\rightarrow P \rightarrow S}} & = & 1 + \epsilon \bm{K}'_{\mathrm{2w}}\cdot[\bm{v}_{\mathrm{P}}(t') - \bm{v}_{\mathrm{S}}(t_{\mathrm{E}}) ] + \epsilon \bm{K}''_{\mathrm{2w}}\cdot[\bm{v}_{\mathrm{S}}(t_{\mathrm{R}}) - \bm{v}_{\mathrm{P}}(t') ]\nonumber\\
  & & + \epsilon^2 \{ \bm{K}'_{\mathrm{2w}} \cdot [\bm{v}_{\mathrm{P}}(t') - \bm{v}_{\mathrm{S}}(t_{\mathrm{E}}) ] \} \{ \bm{K}''_{\mathrm{2w}} \cdot[\bm{v}_{\mathrm{S}}(t_{\mathrm{R}}) - \bm{v}_{\mathrm{P}}(t') ] \}\nonumber\\
  & & -\epsilon^2 [\bm{K}'_{\mathrm{2w}} \cdot\bm{v}_{\mathrm{P}}(t')][\bm{K}'_{\mathrm{2w}}\cdot\bm{v}_{\mathrm{S}}(t_{\mathrm{E}})] + \epsilon^2 [\bm{K}'_{\mathrm{2w}} \cdot\bm{v}_{\mathrm{S}}(t_{\mathrm{E}})]^2\nonumber\\
  & & -\epsilon^2 [\bm{K}''_{\mathrm{2w}}\cdot\bm{v}_{\mathrm{S}}(t_{\mathrm{R}})][\bm{K}''_{\mathrm{2w}} \cdot\bm{v}_{\mathrm{P}}(t')] + \epsilon^2 [\bm{K}''_{\mathrm{2w}} \cdot\bm{v}_{\mathrm{P}}(t')]^2\nonumber\\
  & & +\frac{1}{2}\epsilon^2 [\bm{v}^2_{\mathrm{P}}(t') - \bm{v}^2_{\mathrm{S}}(t_{\mathrm{E}})] + \epsilon^2 \bigg\{\sum_{A} U_A [\bm{y}_{\mathrm{P}}(t')] - \sum_{A} U_A [\bm{y}_{\mathrm{S}}(t_{\mathrm{E}})]\bigg\}\nonumber\\
  & & +\frac{1}{2}\epsilon^2 [\bm{v}^2_{\mathrm{S}}(t_{\mathrm{R}}) - \bm{v}^2_{\mathrm{P}}(t')] + \epsilon^2 \bigg\{\sum_{A} U_A [\bm{y}_{\mathrm{S}}(t_{\mathrm{R}})] - \sum_{A} U_A [\bm{y}_{\mathrm{P}}(t')]\bigg\}\nonumber\\
  & & +\frac{1}{2}\epsilon^2 [ \bar{\beta}_{\mathrm{P}} \bm{v}^2_{\mathrm{P}}(t') - \bar{\beta}_{\mathrm{S}} \bm{v}^2_{\mathrm{S}}(t_{\mathrm{E}})] + \epsilon^2 \bigg\{ \sum_{A} \bar{\alpha}^A_{\mathrm{P}} U_A [\bm{y}_{\mathrm{P}}(t')] - \sum_{A} \bar{\alpha}^A_{\mathrm{S}} U_A [\bm{y}_{\mathrm{S}}(t_{\mathrm{E}})]\bigg\}\nonumber\\
  & & +\frac{1}{2}\epsilon^2 [ \bar{\beta}_{\mathrm{S}} \bm{v}^2_{\mathrm{S}}(t_{\mathrm{R}}) - \bar{\beta}_{\mathrm{P}} \bm{v}^2_{\mathrm{P}}(t')] + \epsilon^2 \bigg\{\sum_{A} \bar{\alpha}^A_{\mathrm{S}} U_A [\bm{y}_{\mathrm{S}}(t_{\mathrm{R}})] - \sum_{A} \bar{\alpha}^A_{\mathrm{P}} U_A [\bm{y}_{\mathrm{P}}(t')]\bigg\}\nonumber\\
  & & + \mathcal{O}(\epsilon^3),
\end{eqnarray}
where
\begin{equation}
  \label{}
  \bm{K}'_{\mathrm{2w}} = - \frac{\bm{y}_{\mathrm{P}}(t')-\bm{y}_{\mathrm{S}}(t_{\mathrm{E}})}{|\bm{y}_{\mathrm{P}}(t')-\bm{y}_{\mathrm{S}}(t_{\mathrm{E}})|} \quad \mathrm{and} \quad  \bm{K}''_{\mathrm{2w}} = - \frac{\bm{y}_{\mathrm{S}}(t_{\mathrm{R}})-\bm{y}_{\mathrm{P}}(t')}{|\bm{y}_{\mathrm{S}}(t_{\mathrm{R}})-\bm{y}_{\mathrm{P}}(t')|}.
\end{equation}

In a special case that $t_{\mathrm{E}}=t'=t_{\mathrm{R}}$ so that we omit them, we can have
\begin{equation}
  \label{}
  \frac{\nu_{\mathrm{R}}}{q\nu_{\mathrm{E}}}\bigg|^{t_{\mathrm{E}}=t'=t_{\mathrm{R}}}_\mathrm{S\rightarrow P \rightarrow S}  =  1 - 2 \epsilon  \bm{n}_{\mathrm{PS}}\cdot \bm{v}_{\mathrm{PS}} + 2\epsilon^2 ( \bm{n}_{\mathrm{PS}}\cdot \bm{v}_{\mathrm{PS}} )^2  + \mathcal{O}(\epsilon^3),
\end{equation}
where $\bm{v}_{\mathrm{PS}}=\bm{v}_{\mathrm{P}}-\bm{v}_{\mathrm{S}}$, $\bm{n}_{\mathrm{PS}} = \bm{R}_{\mathrm{PS}}/R_{\mathrm{PS}}$, $\bm{R}_{\mathrm{PS}} \equiv \bm{y}_{\mathrm{P}} - \bm{y}_{\mathrm{S}}$ and $ R_{\mathrm{PS}} = |\bm{R}_{\mathrm{PS}}|$. When velocities of the spacecraft and the station are very small, this instantaneous approximation of equality of three times in the above equation can be taken. The condition of $t_{\mathrm{E}}=t'=t_{\mathrm{R}}$ also means the light-time (see next Section for details) is not taken into account.

\subsection{Three-Way Doppler}

In the three-way Doppler, there are two stations. Station 1 ($\mathrm{S}_1$) emits a signal and Station 2 ($\mathrm{S}_2$) receives the signal transmitted by a spacecraft (P). In this open-loop, the shift of the frequency is
\begin{equation}
  \label{threewaydopplerLLILPIg}
  \frac{\nu_{\mathrm{R}}}{q\nu_{\mathrm{E}}}\bigg|_{\mathrm{S_1\rightarrow P \rightarrow S_2}} = \frac{\nu'}{\nu_{\mathrm{E}}} \cdot \frac{\nu_{\mathrm{R}}}{q\nu'} = \mathcal{F}_{\mathrm{S_1\rightarrow P}}(t_{\mathrm{E}},t') \cdot \mathcal{F}_{\mathrm{P\rightarrow S_2}}(t',t_{\mathrm{R}})+ \mathcal{O}(\epsilon^3),
\end{equation}
whose explicit form can be written as
\begin{eqnarray}
  \label{threewaydopplerLLILPIex}
  \frac{\nu_{\mathrm{R}}}{q\nu_{\mathrm{E}}} \bigg|_{\mathrm{S_1\rightarrow P \rightarrow S_2}} & = & 1 + \epsilon \bm{K}'_{\mathrm{3w}}\cdot[\bm{v}_{\mathrm{P}}(t') - \bm{v}_{\mathrm{S}_1}(t_{\mathrm{E}}) ] + \epsilon \bm{K}''_{\mathrm{3w}}\cdot[\bm{v}_{\mathrm{S_2}}(t_{\mathrm{R}}) - \bm{v}_{\mathrm{P}}(t') ]\nonumber\\
  & & + \epsilon^2 \{ \bm{K}'_{\mathrm{3w}} \cdot [\bm{v}_{\mathrm{P}}(t') - \bm{v}_{\mathrm{S}_1}(t_{\mathrm{E}}) ] \} \{ \bm{K}''_{\mathrm{3w}} \cdot[\bm{v}_{\mathrm{S}_2}(t_{\mathrm{R}}) - \bm{v}_{\mathrm{P}}(t') ] \}\nonumber\\
  & & -\epsilon^2 [\bm{K}'_{\mathrm{3w}} \cdot\bm{v}_{\mathrm{P}}(t')][\bm{K}'_{\mathrm{3w}}\cdot\bm{v}_{\mathrm{S}_1}(t_{\mathrm{E}})] + \epsilon^2 [\bm{K}'_{\mathrm{3w}} \cdot\bm{v}_{\mathrm{S}_1}(t_{\mathrm{E}})]^2\nonumber\\
  & & -\epsilon^2 [\bm{K}''_{\mathrm{3w}}\cdot\bm{v}_{\mathrm{S}_2}(t_{\mathrm{R}})][\bm{K}''_{\mathrm{3w}} \cdot\bm{v}_{\mathrm{P}}(t')] + \epsilon^2 [\bm{K}''_{\mathrm{3w}} \cdot\bm{v}_{\mathrm{P}}(t')]^2\nonumber\\
  & & +\frac{1}{2}\epsilon^2 [\bm{v}^2_{\mathrm{P}}(t') - \bm{v}^2_{\mathrm{S}_1}(t_{\mathrm{E}})] + \epsilon^2 \bigg\{\sum_{A} U_A [\bm{y}_{\mathrm{P}}(t')] - \sum_{A} U_A [\bm{y}_{\mathrm{S}_1}(t_{\mathrm{E}})]\bigg\}\nonumber\\
  & & +\frac{1}{2}\epsilon^2 [\bm{v}^2_{\mathrm{S}_2}(t_{\mathrm{R}}) - \bm{v}^2_{\mathrm{P}}(t')] + \epsilon^2 \bigg\{\sum_{A} U_A [\bm{y}_{\mathrm{S}_2}(t_{\mathrm{R}})] - \sum_{A} U_A [\bm{y}_{\mathrm{P}}(t')]\bigg\}\nonumber\\
  & & +\frac{1}{2}\epsilon^2 [ \bar{\beta}_{\mathrm{P}} \bm{v}^2_{\mathrm{P}}(t') - \bar{\beta}_{\mathrm{S}_1} \bm{v}^2_{\mathrm{S}_1}(t_{\mathrm{E}})] + \epsilon^2 \bigg\{ \sum_{A} \bar{\alpha}^A_{\mathrm{P}} U_A [\bm{y}_{\mathrm{P}}(t')] - \sum_{A} \bar{\alpha}^A_{\mathrm{S}_1} U_A [\bm{y}_{\mathrm{S}_1}(t_{\mathrm{E}})]\bigg\}\nonumber\\
  & & +\frac{1}{2}\epsilon^2 [ \bar{\beta}_{\mathrm{S}_2} \bm{v}^2_{\mathrm{S}_2}(t_{\mathrm{R}}) - \bar{\beta}_{\mathrm{P}} \bm{v}^2_{\mathrm{P}}(t')] + \epsilon^2 \bigg\{\sum_{A} \bar{\alpha}^A_{\mathrm{S}_2} U_A [\bm{y}_{\mathrm{S}_2}(t_{\mathrm{R}})] - \sum_{A} \bar{\alpha}^A_{\mathrm{P}} U_A [\bm{y}_{\mathrm{P}}(t')]\bigg\}\nonumber\\
  & & + \mathcal{O}(\epsilon^3),
\end{eqnarray}
where
\begin{equation}
  \label{}
  \bm{K}'_{\mathrm{3w}} = - \frac{\bm{y}_{\mathrm{P}}(t')-\bm{y}_{\mathrm{S}_1}(t_{\mathrm{E}})}{|\bm{y}_{\mathrm{P}}(t')-\bm{y}_{\mathrm{S}_1}(t_{\mathrm{E}})|} \quad \mathrm{and} \quad  \bm{K}''_{\mathrm{3w}} = - \frac{\bm{y}_{\mathrm{S}_2}(t_{\mathrm{R}})-\bm{y}_{\mathrm{P}}(t')}{|\bm{y}_{\mathrm{S}_2}(t_{\mathrm{R}})-\bm{y}_{\mathrm{P}}(t')|}.
\end{equation}
In the special case that $t_{\mathrm{E}}=t'=t_{\mathrm{R}}$, we can have
\begin{eqnarray}
  \label{}
  \frac{\nu_{\mathrm{R}}}{q\nu_{\mathrm{E}}} \bigg|^{t_{\mathrm{E}}=t'=t_{\mathrm{R}}}_{\mathrm{S_1\rightarrow P \rightarrow S_2}} & = & 1 - \epsilon \bm{n}_{\mathrm{PS_1}} \cdot \bm{v}_{\mathrm{PS_1}}- \epsilon \bm{n}_{\mathrm{PS_2}} \cdot \bm{v}_{\mathrm{PS_2}} + \epsilon^2 (\bm{n}_{\mathrm{PS_1}} \cdot \bm{v}_{\mathrm{PS_1}})(\bm{n}_{\mathrm{PS_2}} \cdot \bm{v}_{\mathrm{PS_2}})\nonumber\\
  & & -\epsilon^2 (\bm{n}_{\mathrm{PS_1}} \cdot\bm{v}_{\mathrm{P}})(\bm{n}_{\mathrm{PS_1}} \cdot\bm{v}_{\mathrm{S}_1}) + \epsilon^2 (\bm{n}_{\mathrm{PS_1}} \cdot\bm{v}_{\mathrm{S}_1})^2\nonumber\\
  & & -\epsilon^2 (\bm{n}_{\mathrm{PS_2}} \cdot\bm{v}_{\mathrm{P}})(\bm{n}_{\mathrm{PS_2}}\cdot\bm{v}_{\mathrm{S}_2}) + \epsilon^2 (\bm{n}_{\mathrm{PS_2}} \cdot\bm{v}_{\mathrm{P}})^2\nonumber\\
  & & +\frac{1}{2}\epsilon^2 ( \bm{v}^2_{\mathrm{S}_2}- \bm{v}^2_{\mathrm{S}_1} ) + \epsilon^2 \bigg[ \sum_{A} U_A (\bm{y}_{\mathrm{S}_2})- \sum_{A} U_A (\bm{y}_{\mathrm{S}_1})\bigg]\nonumber\\
  & & +\frac{1}{2}\epsilon^2 ( \bar{\beta}_{\mathrm{S}_2} \bm{v}^2_{\mathrm{S}_2} - \bar{\beta}_{\mathrm{S}_1} \bm{v}^2_{\mathrm{S}_1}) + \epsilon^2 \bigg[ \sum_{A} \bar{\alpha}^A_{\mathrm{S}_2} U_A (\bm{y}_{\mathrm{S}_2})  - \sum_{A} \bar{\alpha}^A_{\mathrm{S}_1} U_A (\bm{y}_{\mathrm{S}_1}) \bigg]\nonumber\\
  & & + \mathcal{O}(\epsilon^3),
\end{eqnarray}
where $\bm{v}_{\mathrm{PS_{1/2}}}=\bm{v}_{\mathrm{P}}-\bm{v}_{\mathrm{S}_{1/2}}$, $\bm{n}_{\mathrm{PS}_{1/2}} = \bm{R}_{\mathrm{PS}_{1/2}}/R_{\mathrm{PS}_{1/2}}$, $\bm{R}_{\mathrm{PS}_{1/2}} \equiv \bm{y}_{\mathrm{P}} - \bm{y}_{\mathrm{S}_{1/2}}$ and $ R_{\mathrm{PS}_{1/2}} = |\bm{R}_{\mathrm{PS}_{1/2}}|$. This equation can go back to the Eq. (28) in \citet{Cao2011JoA32.1583} when LLI and LPI are valid.

Although these theoretical models have been established [see Equations (\ref{onewaydopplerLLILPIg}), (\ref{twowaydopplerLLILPIg}) and (\ref{threewaydopplerLLILPIg})], they are still difficult to practice because of their dependence on $t_{\mathrm{E}}$ and/or $t'$ which are usually unavailable in real measurements. In order to solve this problem and make these models depending on the time of reception of the signal $t_{\mathrm{R}}$ only, we need light-time solution \citep{Moyer2000}.

\section{Light-time solution}
\label{sec:lighttime}

The primary contribution of light-time solution (LTS) is to bridge the gaps among $t_{\mathrm{E}}$, $t'$ and $t_{\mathrm{R}}$ \citep[see Chapter 8 in][for details]{Moyer2000}. In a general case, $t_{\mathrm{E}}$ and $t_{\mathrm{R}}$ relate as
\begin{equation}
  \label{}
  \Delta t \equiv (t_{\mathrm{R}}-t_{\mathrm{E}}) = \epsilon |\bm{y}_{\mathrm{R}}(t_{\mathrm{R}}) - \bm{y}_{\mathrm{E}}(t_{\mathrm{E}}) | + \epsilon^3 \Delta \mathcal{T}_{\mathrm{Shapiro}} + \mathcal{O}(\epsilon^5),
\end{equation}
where the second in the right-hand side is the Shapiro time delay caused by the curvature of the spacetime \citep{Shapiro1964PRL13.789}. The Shapiro delay is intensively studied in \citet{Moyer2000}. For light traveling form Jupiter, grazing the surface of the Sun, and arriving at the Earth, its delay due to the Sun is about $10^{-4}$ s. For light traveling from Saturn, grazing the surface of Jupiter, and arriving the Earth, its effect due to Jupiter's mass is $\sim 10^{-7}$ s. For a one-way case that light travels from Saturn, grazes the surface of the Earth and then stops, this delay caused by the mass of the Earth is $\sim 10^{-10}$ s. The magnitudes of such Shapiro time delays are very much less than the time scales of translational and rotational motions of the emitters and receivers of the Doppler tracking links in the solar system so that we can ignore it in the LTS and keep only
\begin{equation}
  \label{}
  \Delta t \equiv (t_{\mathrm{R}}-t_{\mathrm{E}}) = \epsilon |\bm{y}_{\mathrm{R}}(t_{\mathrm{R}}) - \bm{y}_{\mathrm{E}}(t_{\mathrm{E}}) | + \mathcal{O}(\epsilon^3).
\end{equation}

To solve the above equation numerically, one can use the method of iteration. In this work, we prefer to obtain an explicit solution. Since, in the spacecraft Doppler tracking, the time scales of orbital motions of an emitter and a receiver are usually much larger than the time scales of light propagation $\Delta t$, we can do the Taylor expansion as
\begin{equation}
  \label{}
  \bm{y}_{\mathrm{R}}(t_{\mathrm{R}}) = \bm{y}_{\mathrm{R}}(t_{\mathrm{E}}+\Delta t) = \bm{y}_{\mathrm{R}}(t_{\mathrm{E}}) + \bm{v}_{\mathrm{R}}(t_{\mathrm{E}}) \Delta t + \frac{1}{2} \bm{a}_{\mathrm{R}}(t_{\mathrm{E}}) \Delta t^2 +\mathcal{O}(\Delta t^3).
\end{equation}
and
\begin{equation}
  \label{}
  \bm{y}_{\mathrm{E}}(t_{\mathrm{E}}) = \bm{y}_{\mathrm{E}}(t_{\mathrm{R}}-\Delta t) = \bm{y}_{\mathrm{E}}(t_{\mathrm{R}}) - \bm{v}_{\mathrm{E}}(t_{\mathrm{R}}) \Delta t + \frac{1}{2} \bm{a}_{\mathrm{E}}(t_{\mathrm{R}}) \Delta t^2 +\mathcal{O}(\Delta t^3).
\end{equation}
\citet{Moyer2000} argue that the maximum acceleration in the solar system occurs in the region near the Sun ($a\sim 25-274$ m s$^{-2}$) and at the surface of Jupiter ($a\sim25$ m s$^{-2}$). As long as spacecrafts and stations are outside of these regions, \citet{Moyer2000} suggest the acceleration terms in the above two equations can be safely dropped. If we assume all of the Doppler measurements are recorded in terms of $t_{\mathrm{R}}$, a good enough linear approximation of the LTS is
\begin{equation}
  \label{lighttimeyE}
  \bm{y}_{\mathrm{E}}(t_{\mathrm{E}}) =  \bm{y}_{\mathrm{E}}(t_{\mathrm{R}}) - \bm{v}_{\mathrm{E}}(t_{\mathrm{R}}) \Delta t +\mathcal{O}(\Delta t^2),
\end{equation}
and
\begin{equation}
  \label{lighttimedt}
  \Delta t = \epsilon |\bm{y}_{\mathrm{R}}(t_{\mathrm{R}}) - \bm{y}_{\mathrm{E}}(t_{\mathrm{R}})| +\mathcal{O}(\epsilon^3).
\end{equation}

For practical convenience, we will make Doppler models depend on the time of reception of the signal only by using such a linear LTS, which is good enough for most cases \citep{Moyer2000}.

\subsection{One-Way Doppler with LTS}

With Equation (\ref{lighttimedt}), the one-way Doppler can formally written as
\begin{equation}
  \label{}
  \frac{\nu_{\mathrm{R}}}{\nu_{\mathrm{E}}}\bigg|_{\mathrm{E\rightarrow R}} = \mathcal{F}_{\mathrm{E\rightarrow R}}(t_{\mathrm{E}},t_{\mathrm{R}})  =  \mathcal{F}_{\mathrm{E\rightarrow R}}[t_{\mathrm{R}}-\epsilon |\bm{y}_{\mathrm{R}}(t_{\mathrm{R}}) - \bm{y}_{\mathrm{E}}(t_{\mathrm{R}})|,t_{\mathrm{R}}] + \mathcal{O}(\epsilon^3).
\end{equation}
To obtain its explicit expression, we need the expansion of the unit vector $\bm{K}$ [see Equation (\ref{vectorK})] which is
\begin{equation}
  \label{}
  \bm{K} = - \bm{n}_{\mathrm{RE}}(t_{\mathrm{R}}) - \epsilon \{\bm{v}_{\mathrm{E}}(t_{\mathrm{R}}) -  [ \bm{n}_{\mathrm{RE}}(t_{\mathrm{R}})\cdot \bm{v}_{\mathrm{E}}(t_{\mathrm{R}}) ] \bm{n}_{\mathrm{RE}}(t_{\mathrm{R}}) \}  + \mathcal{O}(\epsilon^2),
\end{equation}
where $\bm{n}_{\mathrm{RE}}(t_{\mathrm{R}}) = \bm{R}_{\mathrm{RE}}(t_{\mathrm{R}})/R_{\mathrm{RE}}(t_{\mathrm{R}})$, $\bm{R}_{\mathrm{RE}}(t_{\mathrm{R}}) \equiv \bm{y}_{\mathrm{R}}(t_{\mathrm{R}}) - \bm{y}_{\mathrm{E}}(t_{\mathrm{R}}) $ and $ R_{\mathrm{RE}}(t_{\mathrm{R}}) = |\bm{R}_{\mathrm{RE}}(t_{\mathrm{R}})| $. Thus, the second term in the right-hand side of Equation (\ref{onewaydopplerLLILPI}) can be rewritten as
\begin{eqnarray}
  \label{}
   \bm{K}\cdot[\bm{v}_{\mathrm{R}}(t_{\mathrm{R}}) - \bm{v}_{\mathrm{E}}(t_{\mathrm{E}})]  & = & - \bm{n}_{\mathrm{RE}}(t_{\mathrm{R}})\cdot [\bm{v}_{\mathrm{R}}(t_{\mathrm{R}}) - \bm{v}_{\mathrm{E}}(t_{\mathrm{R}})] -\epsilon \bigg\{ \bm{n}_{\mathrm{RE}}(t_{\mathrm{R}})\cdot \bm{a}_{\mathrm{E}}(t_{\mathrm{R}})R(t_{\mathrm{R}}) \nonumber\\
   & & + \bm{v}_{\mathrm{E}}(t_{\mathrm{R}})\cdot\bm{v}_{\mathrm{R}}(t_{\mathrm{R}}) -  [ \bm{n}_{\mathrm{RE}}(t_{\mathrm{R}})\cdot \bm{v}_{\mathrm{E}}(t_{\mathrm{R}}) ] [\bm{n}_{\mathrm{RE}}(t_{\mathrm{R}})\cdot\bm{v}_{\mathrm{R}}(t_{\mathrm{R}})] \nonumber\\
   & & - \bm{v}^2_{\mathrm{E}}(t_{\mathrm{R}}) +  [ \bm{n}_{\mathrm{RE}}(t_{\mathrm{R}})\cdot \bm{v}_{\mathrm{E}}(t_{\mathrm{R}}) ]^2\bigg\} + \mathcal{O}(\epsilon^2).
\end{eqnarray}
Finally, up to the order of $\epsilon^3$, the shift of the frequency with possible violations LLI and LPI for the one-way Doppler tracking in terms of $t_{\mathrm{R}}$ is
\begin{eqnarray}
  \label{}
  \frac{\nu_{\mathrm{R}}}{\nu_{\mathrm{E}}} \bigg|_{\mathrm{E\rightarrow R}} & = & \mathcal{F}_{\mathrm{E\rightarrow R}}[t_{\mathrm{R}}-\epsilon |\bm{y}_{\mathrm{R}}(t_{\mathrm{R}}) - \bm{y}_{\mathrm{E}}(t_{\mathrm{R}})|,t_{\mathrm{R}}] + \mathcal{O}(\epsilon^3)\nonumber\\
  & = & 1  - \epsilon \bm{n}_{\mathrm{RE}}(t_{\mathrm{R}})\cdot \bm{v}_{\mathrm{RE}}(t_{\mathrm{R}}) -\epsilon^2 \bm{v}_{\mathrm{E}}(t_{\mathrm{R}})\cdot\bm{v}_{\mathrm{R}}(t_{\mathrm{R}}) -\epsilon^2 \bm{n}_{\mathrm{RE}}(t_{\mathrm{R}})\cdot \bm{a}_{\mathrm{E}}(t_{\mathrm{R}})R(t_{\mathrm{R}}) \nonumber\\
   & & +\frac{1}{2}\epsilon^2 \bm{v}^2_{\mathrm{R}}(t_{\mathrm{R}}) + \frac{1}{2}\epsilon^2  \bm{v}^2_{\mathrm{E}}(t_{\mathrm{R}}) + \epsilon^2 \sum_{A}  U_A[\bm{y}_{\mathrm{R}}(t_{\mathrm{R}})] - \epsilon^2 \sum_{A}  U_A [\bm{y}_{\mathrm{E}}(t_{\mathrm{R}})]\nonumber\\
   & & +\frac{1}{2}\epsilon^2 \bar{\beta}_{\mathrm{R}} \bm{v}^2_{\mathrm{R}}(t_{\mathrm{R}}) - \frac{1}{2}\epsilon^2 \bar{\beta}_{\mathrm{E}} \bm{v}^2_{\mathrm{E}}(t_{\mathrm{R}}) + \epsilon^2 \sum_{A} \bar{\alpha}_{\mathrm{R}}^{A} U_A[\bm{y}_{\mathrm{R}}(t_{\mathrm{R}})] - \epsilon^2 \sum_{A} \bar{\alpha}_{\mathrm{E}}^{A} U_A [\bm{y}_{\mathrm{E}}(t_{\mathrm{R}})] \nonumber\\
  & & + \mathcal{O}(\epsilon^3),
\end{eqnarray}
where $\bm{v}_{\mathrm{RE}}=\bm{v}_{\mathrm{R}}-\bm{v}_{\mathrm{E}}$. When $\bm{v}_{\mathrm{R}}=0$, $\bm{a}_{\mathrm{E}}=0$, $U_{A}=0$ and $\bar{\beta}_{\mathrm{R/E}}=\bar{\alpha}_{\mathrm{R/E}}=0$, the above equation can go back to special relativistic transverse Doppler \citep{Landau1975CTFBook}. The possible deviation in the redshift $z$ from the prediction by EEP is
\begin{eqnarray}
  \label{}
  \delta z \bigg|_{\mathrm{E\rightarrow R}} & \equiv & \frac{\nu_{\mathrm{R}}}{q\nu_{\mathrm{E}}}\bigg|_{\mathrm{E\rightarrow R}} - \frac{\nu_{\mathrm{R}}}{q\nu_{\mathrm{E}}}\bigg|^{\mathrm{EEP}}_{\mathrm{E\rightarrow R}} + \mathcal{O}(\epsilon^3) \nonumber\\
  & = &  \frac{1}{2}\epsilon^2 \bar{\beta}_{\mathrm{R}} \bm{v}^2_{\mathrm{R}}(t_{\mathrm{R}}) - \frac{1}{2}\epsilon^2 \bar{\beta}_{\mathrm{E}} \bm{v}^2_{\mathrm{E}}(t_{\mathrm{R}}) + \epsilon^2  \sum_{A} \bar{\alpha}_{\mathrm{R}}^{A} U_A[\bm{y}_{\mathrm{R}}(t_{\mathrm{R}})] - \epsilon^2 \sum_{A} \bar{\alpha}_{\mathrm{E}}^{A} U_A [\bm{y}_{\mathrm{E}}(t_{\mathrm{R}})] \nonumber\\
  & & + \mathcal{O}(\epsilon^3).
\end{eqnarray}

\subsection{Two-Way Doppler with LTS}

In the case of two-way Doppler, after considering LTS, we have
\begin{eqnarray}
  \label{}
  \frac{\nu_{\mathrm{R}}}{q\nu_{\mathrm{E}} }\bigg|_\mathrm{S\rightarrow P \rightarrow S} & = &  \mathcal{F}_{\mathrm{S\rightarrow P}}(t_{\mathrm{E}},t') \cdot \mathcal{F}_{\mathrm{P\rightarrow S}}(t',t_{\mathrm{R}}) + \mathcal{O}(\epsilon^3)\nonumber\\
   & = &  \mathcal{F}_{\mathrm{S\rightarrow P}}[t'-\epsilon |\bm{y}_{\mathrm{P}}(t') - \bm{y}_{\mathrm{S}}(t')|,t'] \cdot \mathcal{F}_{\mathrm{P\rightarrow S}}(t',t_{\mathrm{R}}) + \mathcal{O}(\epsilon^3).
\end{eqnarray}
After substituting $t' = t_{\mathrm{R}}-\epsilon |\bm{y}_{\mathrm{S}}(t_{\mathrm{R}}) - \bm{y}_{\mathrm{P}}(t_{\mathrm{R}})|$ into the above one and expand it with respect to $\epsilon$, we can obtain
\begin{eqnarray}
  \label{twowaydopplerLLILPIexLTS}
  \frac{\nu_{\mathrm{R}}}{q\nu_{\mathrm{E}} }\bigg|_\mathrm{S\rightarrow P \rightarrow S} & = & 1 -\epsilon 2 \bm{n}_{\mathrm{PS}} ( t_{\mathrm{R}}) \cdot \bm{v}_{\mathrm{PS}}(t_{\mathrm{R}}) + 2 \epsilon^2  \bm{v}^2_{\mathrm{PS}}(t_{\mathrm{R}}) \nonumber\\
   & & +2 \epsilon^2  [\bm{n}_{\mathrm{PS}}(t_{\mathrm{R}})\cdot \bm{a}_{\mathrm{PS}}(t_{\mathrm{R}})] R_{\mathrm{PS}}(t_{\mathrm{R}})  + \mathcal{O}(\epsilon^3),
\end{eqnarray}
where $\bm{a}_{\mathrm{PS}} = \bm{a}_{\mathrm{P}} - \bm{a}_{\mathrm{S}}$. Since possible violations of LLI and LPI have opposite signs in the uplink and downlink of the two-way Doppler, they cancel out in this close-loop. It suggests that these violations can not affect two-way Doppler under such linear approximation of LTS [Equations (\ref{lighttimeyE}) and (\ref{lighttimedt})], i.e. $\delta z |^{\mathrm{EEP}}_\mathrm{S\rightarrow P \rightarrow S} = \mathcal{O}(\epsilon^3)$. The effect of a more general approximation of LTS on two-way Doppler will be investigated in our next moves.

\subsection{Three-Way Doppler with LTS}

Applying similar procedure which is applied to one-way and two-way Doppler, we can obtain three-way Doppler with LTS as
\begin{eqnarray}
  \label{}
  \frac{\nu_{\mathrm{R}}}{q\nu_{\mathrm{E}} } \bigg|_\mathrm{S_1\rightarrow P \rightarrow S_2} & = &  \mathcal{F}_{\mathrm{S_1\rightarrow P}}(t_{\mathrm{E}},t') \cdot \mathcal{F}_{\mathrm{P\rightarrow S_2}}(t',t_{\mathrm{R}}) + \mathcal{O}(\epsilon^3)\nonumber\\
   & = &  \mathcal{F}_{\mathrm{S_1\rightarrow P}}[t'-\epsilon |\bm{y}_{\mathrm{P}}(t') - \bm{y}_{\mathrm{S}_1}(t')|,t'] \cdot \mathcal{F}_{\mathrm{P\rightarrow S_2}}(t',t_{\mathrm{R}}) + \mathcal{O}(\epsilon^3),
\end{eqnarray}
where  $t' = t_{\mathrm{R}}-\epsilon |\bm{y}_{\mathrm{S}_2}(t_{\mathrm{R}}) - \bm{y}_{\mathrm{P}}(t_{\mathrm{R}})|$. After Taylor expansion with respect to $\epsilon$, we have
\begin{eqnarray}
  \label{threewaydopplerLLILPIexLTS}
  \frac{\nu_{\mathrm{R}}}{q\nu_{\mathrm{E}}} \bigg|_\mathrm{S_1\rightarrow P \rightarrow S_2}  & = &    1  -\epsilon \bigg[ \bm{n}_{\mathrm{PS_1}} ( t_{\mathrm{R}}) \cdot \bm{v}_{\mathrm{PS_1}}(t_{\mathrm{R}}) + \bm{n}_{\mathrm{PS_2}}(t_{\mathrm{R}})\cdot \bm{v}_{\mathrm{P}\mathrm{S}_2}(t_{\mathrm{R}}) \bigg] \nonumber\\
  & & +\epsilon^2 \bigg[ \mathcal{R}^{\mathrm{P}}_{\mathrm{S_2S_1}}(t_{\mathrm{R}}) \bm{v}^2_{\mathrm{PS}_1}( t_{\mathrm{R}})  - \bm{v}_{\mathrm{P}}(t_{\mathrm{R}}) \cdot \bm{v}_{\mathrm{S}_1}(t_{\mathrm{R}}) - \bm{v}_{\mathrm{P}}(t_{\mathrm{R}})\cdot\bm{v}_{\mathrm{S}_2}(t_{\mathrm{R}})  \nonumber\\
  & & \qquad +  \bm{v}^2_{\mathrm{P}}(t_{\mathrm{R}}) + \frac{1}{2}  \bm{v}^2_{\mathrm{S}_1}(t_{\mathrm{R}})+ \frac{1}{2}  \bm{v}^2_{\mathrm{S}_2}(t_{\mathrm{R}}) \bigg]\nonumber\\
  & & + \epsilon^2 \bigg\{ [ \bm{n}_{\mathrm{PS}_1}(t_{\mathrm{R}})\cdot \bm{v}_{\mathrm{P}\mathrm{S}_1}(t_{\mathrm{R}})]  [ \bm{n}_{\mathrm{PS_2}}(t_{\mathrm{R}})\cdot \bm{v}_{\mathrm{P}\mathrm{S}_2}(t_{\mathrm{R}})]  \nonumber\\
  & & \qquad - \mathcal{R}^{\mathrm{P}}_{\mathrm{S_2S_1}}(t_{\mathrm{R}}) [\bm{n}_{\mathrm{PS}_1}( t_{\mathrm{R}}) \cdot \bm{v}_{\mathrm{PS}_1}( t_{\mathrm{R}})]^2 \bigg\} \nonumber\\
  & & +\epsilon^2 \bigg[ \bm{n}_{\mathrm{PS}_1}( t_{\mathrm{R}}) \cdot \bm{a}_{\mathrm{PS}_1}( t_{\mathrm{R}})R_{\mathrm{PS}_2}(t_{\mathrm{R}}) - \bm{n}_{\mathrm{PS}_1}(t_{\mathrm{R}})\cdot \bm{a}_{\mathrm{S}_1}(t_{\mathrm{R}})R_{\mathrm{PS}_1}(t_{\mathrm{R}}) \nonumber\\
   & & \qquad + \bm{n}_{\mathrm{PS_2}}(t_{\mathrm{R}})\cdot \bm{a}_{\mathrm{P}}(t_{\mathrm{R}})R_{\mathrm{PS_2}}(t_{\mathrm{R}}) \bigg] \nonumber\\
  & & + \epsilon^2 \bigg\{ \sum_{A} U_A[\bm{y}_{\mathrm{S}_2}(t_{\mathrm{R}})] -  \sum_{A}  U_A [\bm{y}_{\mathrm{S}_1}(t_{\mathrm{R}})] \bigg\}\nonumber\\
  & & +\frac{1}{2}\epsilon^2 \bigg[ \bar{\beta}_{\mathrm{S}_2} \bm{v}^2_{\mathrm{S}_2}(t_{\mathrm{R}}) - \bar{\beta}_{\mathrm{S}_1} \bm{v}^2_{\mathrm{S}_1}(t_{\mathrm{R}})  \bigg]\nonumber\\
  & & + \epsilon^2 \bigg\{ \sum_{A} \bar{\alpha}^A_{\mathrm{S}_2} U_A[\bm{y}_{\mathrm{S}_2}(t_{\mathrm{R}})] -  \sum_{A} \bar{\alpha}^A_{\mathrm{S}_1} U_A [\bm{y}_{\mathrm{S}_1}(t_{\mathrm{R}})] \bigg\}\nonumber\\
  & & + \mathcal{O}(\epsilon^3),
\end{eqnarray}
where $\bm{a}_{\mathrm{PS_{1/2}}}=\bm{a}_{\mathrm{P}}-\bm{a}_{\mathrm{S}_{1/2}}$ and 
\begin{equation}
  \label{}
  \mathcal{R}^{\mathrm{P}}_{\mathrm{S_2S_1}}(t_{\mathrm{R}}) \equiv \frac{R_{\mathrm{PS}_2}(t_{\mathrm{R}})}{R_{\mathrm{PS}_1}(t_{\mathrm{R}})}.
\end{equation}

The possibly resulting deviation in the redshift $z$ from the prediction by EEP is
\begin{eqnarray}
  \label{}
  \delta z \bigg|_\mathrm{S_1 \rightarrow P \rightarrow S_2} & \equiv & \frac{\nu_{\mathrm{R}}}{q\nu_{\mathrm{E}}}\bigg|_{\mathrm{S_1 \rightarrow P \rightarrow S_2}} - \frac{\nu_{\mathrm{R}}}{q\nu_{\mathrm{E}}}\bigg|^{\mathrm{EEP}}_{\mathrm{S_1 \rightarrow P \rightarrow S_2}} + \mathcal{O}(\epsilon^3) \nonumber\\
  & = &  \frac{1}{2}\epsilon^2 \bigg[ \bar{\beta}_{\mathrm{S}_2} \bm{v}^2_{\mathrm{S}_2}(t_{\mathrm{R}}) - \bar{\beta}_{\mathrm{S}_1} \bm{v}^2_{\mathrm{S}_1}(t_{\mathrm{R}})  \bigg] + \epsilon^2 \bigg\{ \sum_{A} \bar{\alpha}^A_{\mathrm{S}_2} U_A[\bm{y}_{\mathrm{S}_2}(t_{\mathrm{R}})] -  \sum_{A} \bar{\alpha}^A_{\mathrm{S}_1} U_A [\bm{y}_{\mathrm{S}_1}(t_{\mathrm{R}})] \bigg\}\nonumber\\
  & & + \mathcal{O}(\epsilon^3).
\end{eqnarray}
It indicates that possible violations of LLI and LPI associate with two stations only in three-way Doppler so that it is better to set the stations at places with significant differences in velocities and gravitational potentials to obtain high level of sensitivity for tests. In order to discuss the possibility of detection, we consider a special and optimistic case here as the first step: the stations $\mathrm{S}_1$ and $\mathrm{S}_2$ are two ships respectively located the north pole and the equator of the Earth; the gravitational potential of the Sun is taken into account only; and it is assumed like a sub-case in \citet{Krisher1993PRL70.2213} that $\bar{\beta}_{\mathrm{S}_1} = \bar{\beta}_{\mathrm{S}_2} = \bar{\beta} \sim 10^{-2}$ and ${\alpha}^{\odot}_{\mathrm{S}_1} = {\alpha}^{\odot}_{\mathrm{S}_2} = \bar{\alpha} \sim 10^{-2}$. Then we can have
\begin{eqnarray}
  \label{}
  \delta z \bigg|_\mathrm{S_1 \rightarrow P \rightarrow S_2} & = &  \frac{1}{2}\epsilon^2 \bar{\beta} \bigg[ \bm{v}^2_{\mathrm{S}_2}(t_{\mathrm{R}}) - \bm{v}^2_{\mathrm{S}_1}(t_{\mathrm{R}})  \bigg] + \epsilon^2 \bar{\alpha} \bigg\{ \sum_{A} U_A[\bm{y}_{\mathrm{S}_2}(t_{\mathrm{R}})] -  \sum_{A}  U_A [\bm{y}_{\mathrm{S}_1}(t_{\mathrm{R}})] \bigg\}\nonumber\\
  & \sim & 10^{-12},
\end{eqnarray}
which also yields $\delta v = c \,\delta z|_\mathrm{S_1 \rightarrow P \rightarrow S_2} \sim 3 \times 10^{-4}$ m s$^{-1}$. Although this magnitude of $\delta z|_\mathrm{S_1 \rightarrow P \rightarrow S_2}$ may be able to detect with current stage of Doppler tracking, the configuration of the stations are too particular. In our next moves, we will focus on case studies of some experiments conducted with real facilities.

\section{Conclusions and Discussion}

\label{sec:con}

Currently, two-way and three-way spacecraft Doppler tracking techniques are widely used and playing important roles in control and navigation for deep space missions. Starting from one-way Doppler model \citep{Krisher1993PRL70.2213}, we extend the models of two-way and three-way Doppler by making them [see Equations (\ref{twowaydopplerLLILPIex}) and (\ref{threewaydopplerLLILPIex})] include possible violations of LLI and LPI in order to test EEP which is the cornerstone of GR and all other metric theories of gravity \citep{Will1993TEGP,Will2006LRR9.3}. After taking the finite speed of light into account, which is so-called light-time solution (LTS) \citep{Moyer2000}, we have these models depending on the time of reception of the signal only for practical convenience [see Equations (\ref{twowaydopplerLLILPIexLTS}) and (\ref{threewaydopplerLLILPIexLTS})]. We find that possible violations of LLI and LPI can not affect two-way Doppler under linear approximation of LTS [Equations (\ref{lighttimeyE}) and (\ref{lighttimedt})] although this approximation is sufficiently good for most cases in the solar system \citep{Moyer2000}. We also show that, in three-way Doppler, possible violations of LLI and LPI associate with two stations only, which suggests that it is better to set the stations at places with significant differences in velocities and gravitational potentials to obtain high level of sensitivity for tests. 

In practice, Doppler measurements certainly suffer various noise, such as frequency standard noise, plasma scintillation noise, tropospheric scintillation noise, antenna mechanical noise, ground electronics noise, spacecraft transponder noise, thermal noise in the ground and spacecraft receivers, and spacecraft unmodeled motion \citep[see][for a review]{Armstrong2006LRR9.1}. Although studies on these noise are out of the scope of this paper, they are extremely important for a positive detection. In our next moves, we will focus on case studies of some specific missions.

\normalem
\begin{acknowledgements}
The work of X.-M.D. is funded by the Natural Science Foundation of China (Grant No. 11103085) and the Fundamental Research Program of Jiangsu Province of China (Grant No. BK20131461). The work of Y.X. is supported by the National Natural Science Foundation of China (Grant No. 11103010), the Fundamental Research Program of Jiangsu Province of China (Grant No. BK2011553) and the Research Fund for the Doctoral Program of Higher Education of China (Grant No. 20110091120003).

\end{acknowledgements}

\bibliographystyle{raa.bst}
\bibliography{ms1617.bib}

\end{document}